# Threshold Behaviour in Light Reflection Tuning the Disorder in Photonic Media

Michele Bellingeri[a] and Francesco Scotognella*[b]

[a] *Dipartimento di Scienze Ambientali, Università di Parma, Parco Area delle Scienze, 33/A 43100 Parma, Italy*

[b] *Dipartimento di Fisica, Politecnico di Milano, piazza Leonardo da Vinci 32, 20133 Milano, Italy.*
*To whom correspondence should be addressed: francesco.scotognella@polimi.it

**Abstract**

**The optical properties of materials are strongly influenced by disorder. Control of disorder in photonic materials can unveil interesting optical properties. We have found an engineered photonic structure for which the average light reflection shows a linear behaviour with a slope change in a broad range of wavelengths. Such change in slope is due to a specific degree of disorder, which is quantified by the Shannon index.**

## Introduction

In the last years, the study of the propagation of light in complex dielectric materials has attracted increasing attention, since they are extremely widespread in nature. However, a comprehensive understanding of their behaviour has not been achieved yet. A distinction can be made among these materials, in which fluctuations of the refractive index occur at the order of the wavelength of light, by considering their grade of order. Completely ordered dielectric materials are called photonic crystals. Between photonic crystals and photons there are analogous relations that exist between semiconductor crystals and electrons: for a certain range of energies and certain wave vectors, light is not allowed to propagate through the medium [1-4]. Such peculiar structures are naturally occurring. Otherwise they can be fabricated with several techniques having a periodicity in one-, two- and three-dimensions [5-9]. On the other side, completely disordered materials, i.e. opaque materials, follow the Anderson localization, that describes how the introduction of disorder can lead from a conducting to an insulating phase of the material [10].

However, there are materials which represent an intermediate phase between fully periodic and fully disordered media. A very interesting class of materials are quasicrystals, which do not have a unit cell and translation symmetry, but they have a rotational symmetry (noncrystallographic), long-range order, and they display Bragg diffraction [11-12]. Many of the properties of quasicrystals are nowadays well understood, even though some fundamental questions still remain unanswered. Among such questions, one of the most important is about transport in quasicrystals. Recently, the group of Prof. Segev has experimentally observed an enhancement of the transport in quasicrystals by introducing disorder. This work is the first study of transport in quasicrystals in the presence of disorder [13]. Moreover, concepts and methods widely used in statistics have been successfully applied to explain light transport phenomena in materials where the local density of light scattering elements is position-dependent (made by Titanium dioxide particles and a non-trivial size distribution of glass microspheres in a glass matrix) [14-17].

Another way to explain the disorder in a crystal is to describe the non-homogeneity of such crystal. Thus, the homogeneity of a crystal can be quantified by a well known diversity index, the Shannon index [21,22]. It has been demonstrated that it is possible to correlate the average light transmission over a large range of wavelengths to the Shannon index [18-19]. The average light transmission decreases linearly by decreasing the homogeneity (i.e. the Shannon index) of a two-dimensional photonic crystal [18]. Furthermore, in Ref. [19] it has been shown the average light reflection, as a function of the sample length, increases sub-linear in an ideal two-dimensional photonic crystal while increases linear in a non-homogeneous photonic media (with Shannon index $H' = 0.7$).

In Bellingeri et al. [18] average light transmission is investigated in non-homogeneous crystals realized aggregating pillars in clusters of equal size. Now, it is necessary to describe the average transmission in crystals of increasing disorder, but with random cluster size distribution.

To this end, in this paper we study the light reflection as a function of the Shannon index, used to quantify the degree of disorder, from a non-trivial engineered two-dimensional disordered photonic media. Random disordered structures are realized using a fitness model [20]. We have found out that such average reflection shows a pronounced change in slope with disorder corresponding to $0.85 < H' < 0.9$. Moreover, we have observed that the standard deviation of the average reflection is inversely proportional to the Shannon index, i.e. proportional to the disorder in the photonic structure.

**Methods**

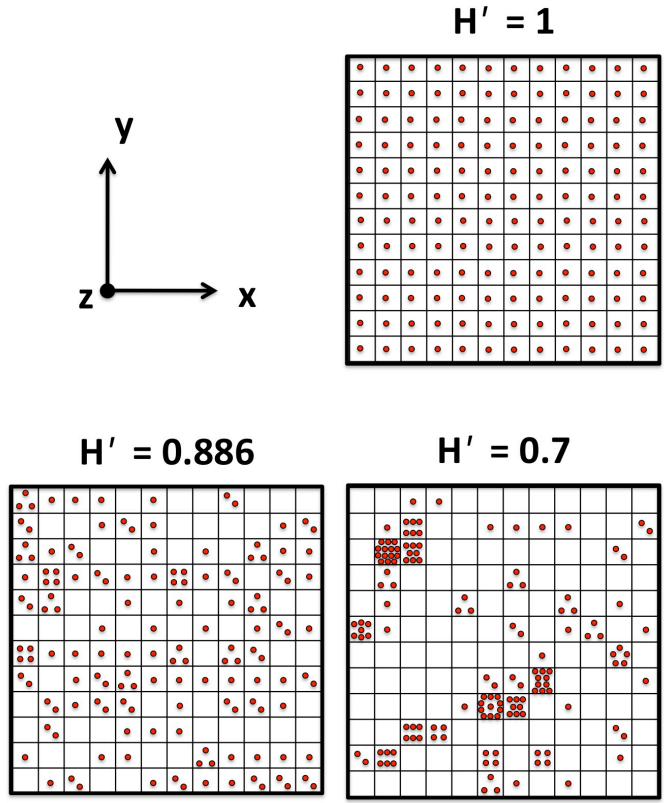

**Figure 1.** 12x12 unit cell photonic crystal (corresponding to a Shannon index $H' = 1$), made by Titanium dioxide (TiO$_2$) and Silicon dioxide (SiO$_2$), and 12x12 unit cell disordered photonic structure corresponding to $H' = 0.886$ and 0.7.

In this work we have considered an ideal two-dimensional photonic crystal [4]. Such photonic structure is a square lattice of dielectric circular pillars, where the pillars have a diameter $d$ of 75 nm and are made of Titanium dioxide. The lattice constant $a$ of the crystal is 300 nm and the matrix in which the pillars are embedded is Silicon dioxide. The refractive indexes of TiO$_2$ and SiO$_2$ are $n_T = 2.45$ and $n_S = 1.46$, respectively. It is remarkable that $n_T d \sim n_S(a-d)$ is satisfied for such a geometrical setting [4]. We consider a 12x12 photonic crystal unit cell with a size of 3.6×3.6 μm. In order to analyze the light transmission as a function of the topological disorder, we have built photonic structures of various grade of homogeneity. The photonic crystal is the most homogeneous one in which each cell contains one pillar. Starting from this, we have built more disordered media by a fitness model [20]. To sum up, in order to realize such disordered structures, we have assigned pillars in cells with a different cell probability or cell fitness [19-20]. It is possible to aggregate pillars in clusters and to build structures of different grade of disorder by tuning cell fitness. The media resulting from this process are characterized by a skewed clusters size without benchmark distribution. Thus, we have a random structure set with an increasing skewness of the clusters size distribution (see Supplementary Information). The most homogeneous, an intermediate, and the most disordered structures are represented in Figure 1. Each 12x12 crystal unit represented in Figure 1 is repeated seven times.

It is possible to correlate the distribution of the pillars in the structure to the Shannon-Wiener index [21,22]. The Shannon-Wiener $H'$ index is a diversity index widely used in statistics and information theory. Technically, such index is a measurement of the information entropy of the distribution and it defined as

$$H' = -\sum_{j=1}^{s} p_j \log p_j \qquad (1)$$

where $p_j$ is the proportion of the $j$-fold species and $s$ is the number of the species. For a certain number of species (parameter $s$ in equation 1), $log(s)$ represents the maximum value of the Shannon index and it corresponds to the maximum topological evenness. Thus, dividing equation (1) by $log(s)$ we constrain $H'$ in (0,1). We use this $H'$ notation and we call $H'$ normalized Shannon-Wiener index. In the ideal two-dimensional photonic crystal, the normalized Shannon-Wiener index has a value of 1, corresponding to the maximum of the homogeneity. On the contrary, in the fitness model crystals, i.e. in the disordered photonic structures, the value of the normalized Shannon-Wiener index is < 1. This implies that the realized photonic structures are less homogeneous than the ideal photonic crystal.

Five different realizations have been considered for each crystal by allocating the clusters among the available 12×12 cells of the original crystal in a random fashion. In this way, the five realizations of each crystal have the same Shannon index, but they are different in topology (i.e. the cluster size distribution is the same but clusters are allocated at random).

As regards the light transmission calculation (and reflection) of the photonic structures through finite element method, we assumed a TM-polarized field, while we used the scalar equation for the transverse electric field component $E_Z$

$$\left(\partial_x^2 + \partial_y^2\right)E_Z + n^2 k_0^2 E_Z = 0 \qquad (2)$$

where $n$ is the refractive index distribution and $k_0$ is the free space wave number [4,23]. As input field, a plane wave with wave vector $k$ directed along the $x$-axis has been assumed. Scattering boundary conditions in the $y$ direction have been used.

**Results and Discussion**

The whole structure has a size 3.6×3.6 μm, repeated seven times along the $x$-axis, and it is depicted in Figure 1. Also for this structure, the 3.6×3.6 μm composition has been repeated seven times. We have selected this seven-fold repetition of 12×12 unit cells since a very interesting behaviour has been observed: the less homogeneous structure shows a linear behaviour of the average reflection as a function of the sample length, while the photonic crystal with the same number of scattering centres shows a strong sub-linear behaviour [19]. In this work we have considered photonic structures with different Shannon index in the range $0.7 \leq H' \leq 1$, and we have calculated their reflection spectra by means of a finite element method. Since we are analysing structures made with transparent materials in the range 450-1400 nm, we can assume that the absorption of the structure is due only to light reflection. In Figure 2, the black line represents the reflection of the photonic crystal repeated seven times, with the photonic band gap at about 925 nm, according to the Bragg-Snell law [1,4,24]. The photonic band gap can be seen also in the more disordered photonic structure corresponding to a Shannon index $H' = 0.886$ (red line), yet in this case the spectrum shows an higher reflection in all the wavelengths out of the gap. The reflection spectrum is significantly different in the disordered structure with $H' = 0.7$, where the photonic band gap is not present and the variations in reflection decrease with respect to the other structures (from 3 to 1).

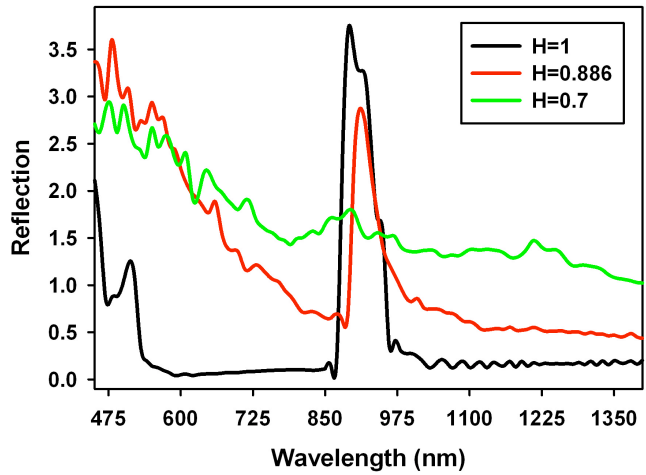

**Figure 2.** Reflection spectra of the two-dimensional 12x12 photonic crystal, corresponding to a Shannon index H'=1 (black line), and reflection spectra of a 12x12 disordered photonic structure corresponding to a Shannon index $H' = 0.886$ (red line) and $H' = 0.886$ (green line).

Figure 3 shows the average reflection in the range 450 – 1400 nm as a function of the Shannon index. We can divide the relationship in two distinct linear trends. The first trend in the range $0.7 \leq H' \leq 0.886$ is described by the linear model with Slope = -143.86 and Intercept = 243.58 ($p < 0.05$). The second trend in the range $0.886 \leq H' \leq 1$ follows the linear model with Slope = -646.43 and Intercept = 694.44 ($p < 0.001$). Thus, we can observe a pronounced slope variation at $H' = 0.886$. The light reflection decreases slightly starting from a crystal with $H' = 0.7$ to $H' = 0.8$, that is in more disordered structures. Once it has reached $H' = 0.886$, the light reflection in more homogeneous structure falls sharply than in the previously disordered range, reaching the minimum in the photonic crystal.

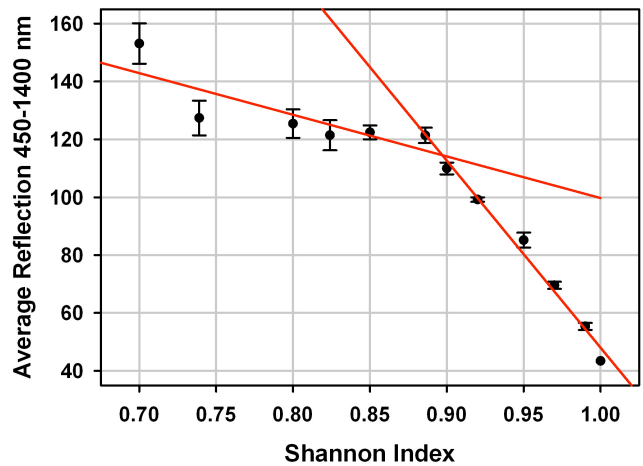

**Figure 3.** Average reflection in the range 450 – 1400 nm as a function of the Shannon index of the Figure S1 in the Supplementary Information shows the average reflection as a function of the Shannon index for different repetitions of 3.6×3.6 µm photonic structure. The behaviour shown in Figure 3 is significantly evident for seven repetitions of the photonic structure.

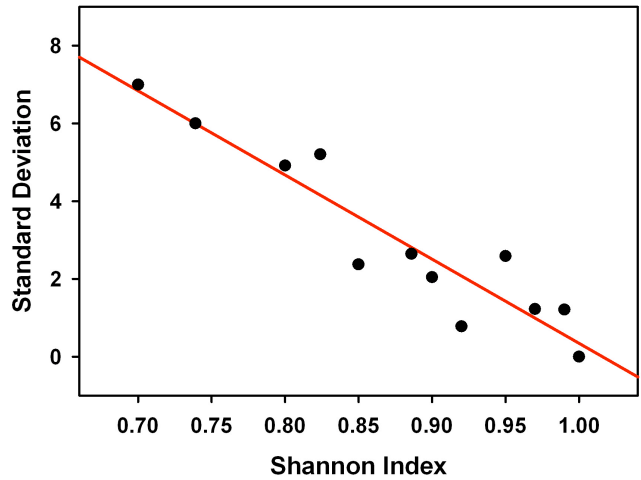

**Figure 4.** Linear model: standard deviation of the average reflection as a function of the Shannon index.

Another remarkable phenomenon correlated to the disorder is the fluctuation of the average reflection by calculating the reflection spectra for the different realizations. In Figure 4, the standard deviation of the average reflection is shown as a function of the Shannon index. Furthermore, we have observed that the more the disorder increases (decrease of the Shannon index) the more the standard deviation of the average reflection increases, indicating that the fluctuation of such average reflection is higher in disordered structures. This trend follows a strong linear decrease with Slope = -21.659 and Intercept = 22 ($p < 0.0001$).

**Conclusion**

The average reflection over the range of wavelengths 450 – 1400 nm has been studied as a function of the disorder in photonic structure with a specific sample length, so that a threshold behaviour in the average reflection is evident: a linear trend occurs with disorder corresponding to a range of Shannon index $0.7 \leq H' \leq 0.886$, up to an abrupt change in slope with another linear behaviour in the range $0.886 \leq H' \leq 1$. Moreover, we have observed that the standard deviation of the average reflection decreases as a function of the Shannon index. This indicates that the bigger the disorder of the photonic structure is, the larger the fluctuation of the average reflection by varying the distribution of pillar clusters in the structure is too. These results represent a step towards a better understanding of the optical properties of photonic structures as a function of their disorder which can be very useful for several applications, such as diagnostic imaging [25] and light harvesting for optoelectronics [26,27].

**Notes and references**

# Supplementary Data

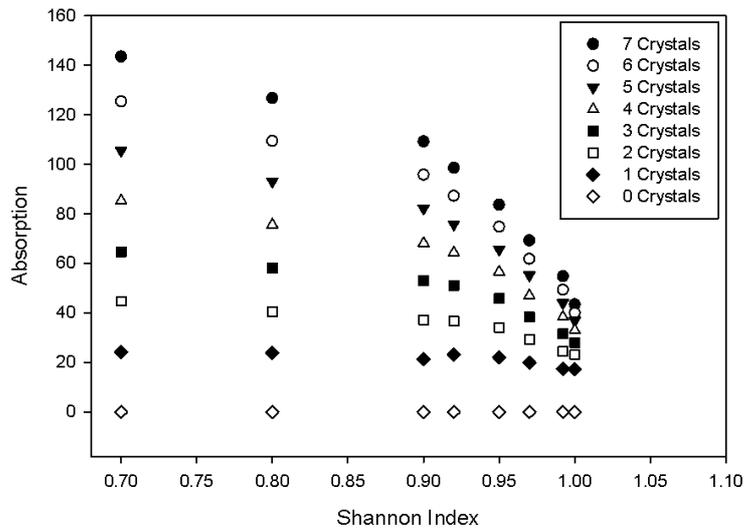

**Figure S1.** Average reflection in the range 450 – 1400 nm as a function of the Shannon index for different repetitions of the photonic structure.

Example of matrices to build fitness model photonic structures: Each element of the matrix is a topological cell of the crystal. The element value indicates the number of scattering centres in that cell.

H'=0.886

| 3 | 1 | 1 | 1 | 0 | 1 | 0 | 0 | 2 | 0 | 0 | 0 |
|---|---|---|---|---|---|---|---|---|---|---|---|
| 2 | 0 | 0 | 1 | 2 | 1 | 0 | 0 | 0 | 0 | 1 | 2 |
| 3 | 1 | 2 | 0 | 0 | 1 | 0 | 1 | 0 | 3 | 1 | 1 |
| 1 | 4 | 1 | 2 | 1 | 1 | 4 | 1 | 2 | 0 | 2 | 1 |
| 2 | 3 | 0 | 0 | 1 | 0 | 1 | 0 | 1 | 3 | 0 | 0 |
| 0 | 0 | 0 | 1 | 0 | 1 | 0 | 1 | 0 | 1 | 2 | 1 |
| 4 | 1 | 1 | 1 | 1 | 1 | 3 | 0 | 3 | 2 | 0 | 0 |
| 2 | 0 | 1 | 2 | 3 | 1 | 1 | 1 | 1 | 1 | 1 | 2 |
| 0 | 2 | 1 | 2 | 2 | 0 | 1 | 0 | 2 | 2 | 1 | 0 |
| 0 | 2 | 0 | 0 | 1 | 1 | 1 | 0 | 0 | 0 | 0 | 0 |
| 1 | 0 | 0 | 1 | 1 | 0 | 0 | 3 | 1 | 1 | 1 | 1 |
| 0 | 1 | 2 | 0 | 0 | 1 | 2 | 1 | 1 | 2 | 2 | 2 |

H'= 0.9215284

| 2 | 2 | 1 | 1 | 1 | 2 | 2 | 1 | 1 | 0 | 0 | 0 |
|---|---|---|---|---|---|---|---|---|---|---|---|
| 0 | 1 | 1 | 0 | 0 | 1 | 1 | 0 | 0 | 0 | 1 | 1 |
| 1 | 2 | 1 | 1 | 1 | 1 | 2 | 1 | 1 | 0 | 1 | 0 |
| 1 | 0 | 0 | 0 | 1 | 0 | 3 | 0 | 1 | 2 | 1 | 0 |
| 2 | 1 | 0 | 0 | 1 | 1 | 1 | 0 | 2 | 1 | 1 | 1 |
| 0 | 2 | 1 | 2 | 1 | 1 | 0 | 0 | 3 | 2 | 1 | 0 |
| 2 | 1 | 3 | 0 | 1 | 3 | 1 | 1 | 1 | 0 | 1 | 1 |
| 1 | 0 | 1 | 0 | 0 | 1 | 2 | 1 | 1 | 2 | 1 | 2 |
| 2 | 2 | 2 | 1 | 2 | 2 | 2 | 1 | 0 | 2 | 1 | 0 |
| 2 | 1 | 0 | 1 | 1 | 1 | 1 | 1 | 2 | 0 | 0 | 2 |
| 1 | 1 | 1 | 0 | 0 | 1 | 2 | 1 | 2 | 0 | 2 | 2 |
| 0 | 1 | 1 | 1 | 2 | 1 | 1 | 1 | 0 | 1 | 1 | 1 |